\documentclass[twocolumn,aps,showpacs,preprintnumbers,lettersize]{revtex4}
\usepackage{amsfonts}
\usepackage{amsmath}
\usepackage{amssymb}
\usepackage{graphicx}

\begin{document}

\title{Electronic Properties of the Hubbard Model on a Frustrated 
       Triangular Lattice}
\author{Bumsoo Kyung}
\affiliation{D\'{e}partement de physique and Regroupement qu\'{e}b\'{e}cois sur les mat%
\'{e}riaux de pointe, Universit\'{e} de Sherbrooke, Sherbrooke, Qu\'{e}bec,
J1K 2R1, Canada}
\date{\today }

\begin{abstract}
   We study 
electronic properties of the Hubbard model on a triangular lattice
using the cellular dynamical mean-field theory.
The interplay of strong geometric frustration and electron correlations
causes a Mott transition at the Hubbard interaction $U/t=10.5$
and an unusual suppression of low energy spin excitations.
Doping of a triangular Mott insulator leads to a quasiparticle peak
(no pseudogap) at the Fermi surface and to an unexpected increase of low energy 
spin excitations, in stark contrast to the unfrustrated square lattice case.
The present results give much insight into strongly frustrated 
electronic systems.
A few predictions are made.
\end{abstract}

\maketitle

   Geometric frustration with strong electronic correlations
is one of the main issues in modern condensed matter physics.
The simplest example is the two dimensional Heisenberg model    
(large $U$ limit of the half-filled Hubbard model)
on a triangular lattice 
in which all three spins cannot be antiparallel at the same time.
The frustrated triangular lattice geometry was argued by  
Anderson~\cite{Anderson:1973}  
to provide an ideal background for the long sought resonating valence 
bond (RVB) state.
The interplay of strong geometric frustration and electronic
correlations is expected to lead to some exotic phases.
Recent discovery of superconductivity in the triangular   
lattice compound Na$_{x}$CoO$_{2}$ $\cdot$ yH$_{2}$0~\cite{Takada:2003}
and a possible quantum spin liquid state in the anisotropic 
triangular lattice organic material 
$\kappa $-(ET)$_{2}$Cu$_{2}$(CN)$_{3}$~\cite{Shimizu:2003}
has further stimulated interest in strongly frustrated electronic systems.

   The electronic properties of the Hubbard and $t-J$ models on a 
triangular lattice have been studied using various analytical and numerical 
techniques. These include  
a high-temperature expansion~\cite{Koretsune:2002}, 
a slave-boson mean-field~\cite{Baskaran:2003,Kumar:2003,Wang:2004},
an RG method~\cite{Honerkamp:2003}, an exact
diagonalization (ED) technique~\cite{Capone:2001,Haerter:2005,Tohyama:2006},
a quantum Monte Carlo (QMC) simulation~\cite{Bulut:2005}, 
a dynamical mean-field theory (DMFT)~\cite{Merino:2005,Aryanpour:2006},
and a variational Monte Carlo approach~\cite{Weber:2006}. 
However, some of the central issues on a triangular lattice  
have not been clearly addressed and still remain an open question:
what is the nature of the ground state of a doped triangular Mott insulator?
is it a non-Fermi liquid (manifested as the presence of a pseudogap
at small doping like on an unfrustrated square lattice)  
or a correlated Fermi liquid? 
what are qualitative differences between the triangular and square 
lattice systems?
In this paper, utilizing recent theoretical progress in quantum cluster 
methods, we address these issues and provide much insight into electronic 
properties in strongly frustrated electronic systems.

   The two-dimensional Hubbard model on a triangular lattice is 
described by 
\begin{equation}
H=\sum_{\langle ij\rangle ,\sigma }t_{ij}c_{i\sigma }^{\dagger }c_{j\sigma
}+U\sum_{i}n_{i\uparrow }n_{i\downarrow }
-\mu \sum_{i\sigma }c_{i\sigma}^{\dagger }c_{i\sigma }
\, ,  \label{eq10}
\end{equation}
where $c_{i\sigma }^{\dagger }$ ($c_{i\sigma }$) are creation (annihilation)
operators for electrons of spin $\sigma $, $n_{i\sigma }=c_{i\sigma
}^{\dagger }c_{i\sigma }$ is the density of $\sigma $ spin electrons,
$U$ is the on-site repulsive interaction,
and $\mu $ is the chemical potential controlling the electron density.

\begin{figure}[t]
\includegraphics[width=8.0cm]{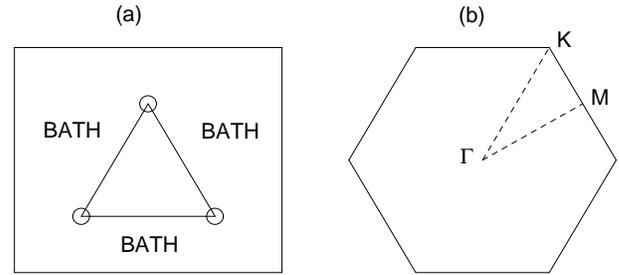}
\caption{(a) Three-site quantum cluster embedded in an effective 
         medium (called bath)
         and (b) first Brillouin zone of the triangular lattice.
        }
\label{cluster.fig}
\end{figure}

   To describe strong electron correlations and geometric frustration 
accurately, we use cellular dynamical mean-field theory 
(CDMFT)~\cite{Kotliar:2001},
a quantum cluster approach that allows one to extend 
DMFT~\cite{GKKR:1996} to incorporate short-range spatial correlations 
explicitly. CDMFT has been benchmarked and is accurate even in one 
dimension~\cite{Bolech:2002,KKT:2006}.
The infinite lattice is tiled with identical clusters
of size $N_{c}$, and the degrees of freedom in the cluster are treated
exactly while the remaining ones are replaced by a bath of non-interacting
electrons that is determined self-consistently (Fig.~\ref{cluster.fig}(a)).
To solve the quantum cluster
embedded in an effective medium, we consider a cluster-bath Hamiltonian
of the form~\cite{Kotliar:2001,Bolech:2002,KKSTCK:2006}
\begin{eqnarray}
H &=&\sum_{\langle \mu \nu \rangle ,\sigma }t_{\mu \nu }c_{\mu \sigma
}^{\dagger }c_{\nu \sigma }+U\sum_{\mu }n_{\mu \uparrow }n_{\mu \downarrow }
+ \sum_{m,\sigma ,\alpha }\varepsilon _{m\sigma }^{\alpha }a_{m\sigma
}^{\dagger \alpha }a_{m\sigma }^{\alpha }
\notag \\
&+&\sum_{m,\mu ,\sigma ,\alpha
}V_{m\mu \sigma }^{\alpha }(a_{m\sigma }^{\dagger \alpha }c_{\mu \sigma }+
\mathrm{H.c.}) \,.  \label{eq20} 
\end{eqnarray}
Here the indices $\mu ,\nu =1,\cdots ,N_{c}$ label sites within the cluster,
and $c_{\mu \sigma }$ and $a_{m\sigma }^{\alpha }$ annihilate electrons on
the cluster and the bath, respectively. $t_{\mu \nu }$ are the hopping matrix
elements within the cluster (the chemical potential $\mu$ is absorbed here),
$\varepsilon _{m\sigma }^{\alpha }$ are the bath
energies and $V_{m\mu \sigma }^{\alpha }$ are the bath-cluster hybridization
matrices.
The exact diagonalization method~\cite{Caffarel:1994} is used to solve the
cluster-bath Hamiltonian Eq.~\ref{eq20} at zero temperature, which has the
advantage of computing dynamical quantities directly in real frequency and
of treating the large $U$ regime without difficulty. 
In the present study we used $N_{c}=3$ sites for the cluster
and $N_{b}=9$ sites for the bath with $m=1,2,3$, $\alpha=1,2,3$.
Although the present
study focuses on a small cluster with additional $9$ bath sites, we
expect our results to be robust with respect to an increase in the cluster
size. This was verified by our recent low (but finite) temperature CDMFT+QMC
calculations~\cite{KKT:2006} where at intermediate to strong coupling a 
4-site cluster accounts for more than $95\%$ of the correlation effect
of the infinite size cluster in the single-particle spectrum. 
Quantitatively similar results were found for different bath sizes 
($N_{b}=6$).
In this paper we focus on the normal state properties 
at and near half-filling.

\begin{figure}[t]
\includegraphics[width=8.0cm]{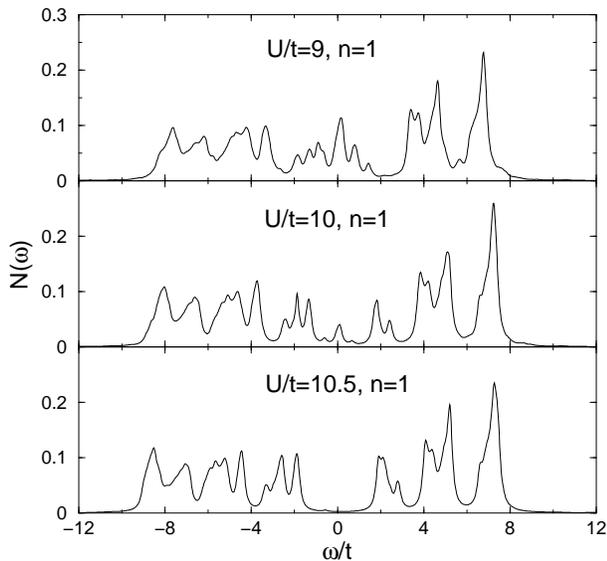}
\caption{Density of states $N(\omega)$ on a triangular lattice 
         at half-filling ($n=1$) with increasing $U/t$.
        }
\label{mott_transition.fig}
\end{figure}

   We first study the Mott transition of the half-filled Hubbard model 
on a triangular lattice. Among several physical quantities one can 
gauge to investigate the Mott transition, we present the density of states
$N(\omega)$ as a function of $U/t$ shown in Fig.~\ref{mott_transition.fig}.
For $U/t < 10.5$ a small spectral weight remains near the Fermi energy
reminiscent of the Kondo resonance observed in DMFT~\cite{GKKR:1996}. 
%
When $U/t$ approaches $10.5$, the spectral weight of the central peak is  
completely pushed away from $\omega=0$ leading to an interaction-driven 
Mott insulator. This critical value of $U/t$ is somewhat smaller than $12$
obtained in the ED study of an isolated 12-site cluster~\cite{Capone:2001} 
and in the DMFT analysis~\cite{Aryanpour:2006}. 
Beyond $U/t=10.5$ it becomes increasingly more difficult to find a convergent 
solution, since the ground state starts to become degenerate.
The way how the Mott transition occurs on a frustrated triangular 
lattice is dramatically different from the unfrustrated square lattice case.
In the latter~\cite{KKSTCK:2006}, with increasing $U/t$ the spectral weight 
always becomes a local minimum at the Fermi energy until it completely vanishes 
near $U/t=6$.
%
\begin{figure}[t]
\includegraphics[width=8.0cm]{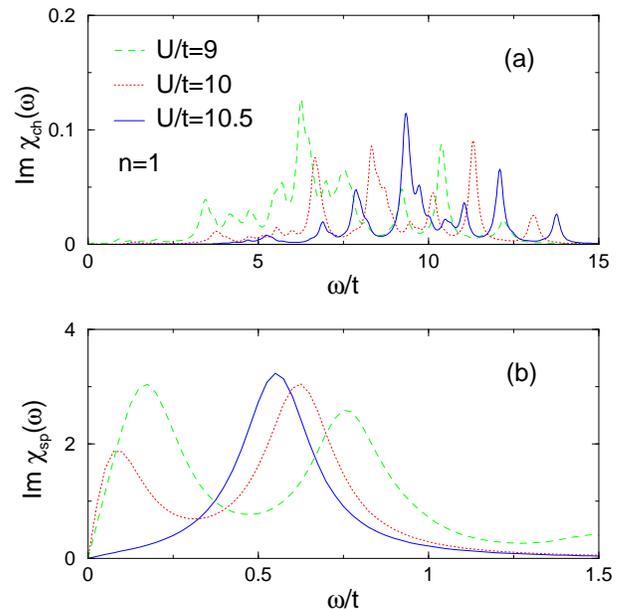}
\caption{(color online) Imaginary part of the local dynamical (a) charge 
         and (b) spin
         correlation functions at half-filling with increasing $U/t$.
         The dashed, dotted and solid curves correspond to 
         $U/t$=9, 10 and 10.5, respectively.
        }
\label{charge_spin.fig}
\end{figure}

   Next we examine the local dynamical charge and spin correlation functions
near the Mott transition
shown in Fig.~\ref{charge_spin.fig}.
With increasing $U/t$ the low energy charge excitations 
are gradually depleted and transfered to high energy near $\omega=U/t$.
On the other hand, the low energy spectrum in the spin correlation function  
undergoes a dramatic change near the Mott transition.
As the Mott transition is approached, the primary low energy excitation
near $w/t=0.2$ moves rapidly to the secondary peak near $w/t=0.5-0.6$
and eventually disappears.
Namely, geometric frustration suppresses the low energy spin excitations,
which is most strongly manifested in the \textit{insulating} state ($U/t=10.5$).
We found similar results for the nearest neighbor charge and spin 
correlation functions.
This feature is in stark contrast to the square lattice case.
In the latter, for the same range of $U/t$ ($9-10.5$)
the low energy spin spectra are almost identical with a sharp primary peak
near $w/t=0.25-0.3$~\cite{comment}  
and a secondary peak near $w/t=0.6-0.7$ similar to 
the case of $U/t=9$ on a triangular lattice.
Near the range of $U/t$ ($4-6$) where a continuous crossover of 
a metal to an insulator occurs
on a square lattice, with increasing $U/t$ the low energy spin spectrum  
near $w/t=0.25$ becomes stronger, which is opposite to the behavior 
found on a triangular lattice.

\begin{figure}[t]
\includegraphics[width=13.5cm,angle=270]{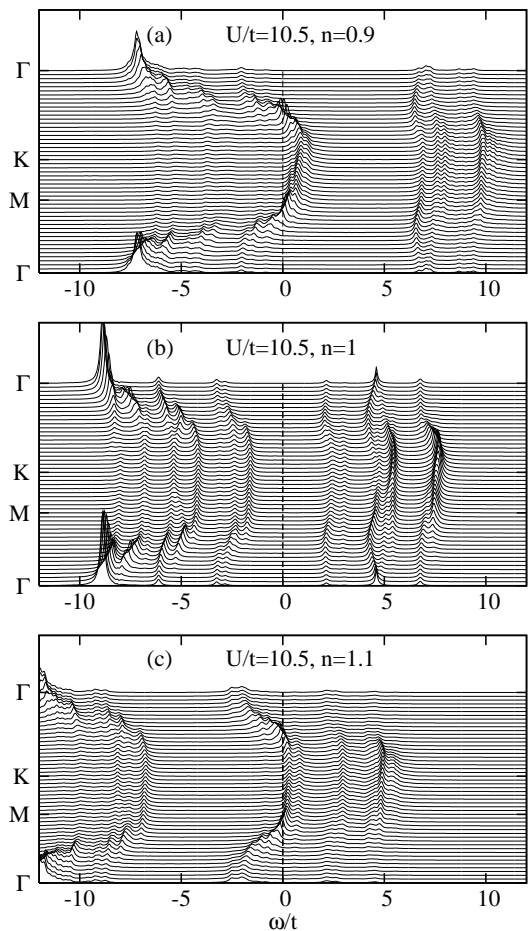}
\caption{Spectral function $A(\vec{k},\omega)$ for (a) $n$=0.9,
         (b) $n$=1, and (c) $n$=1.1 along the path
         $\Gamma$-M-K-$\Gamma$ for $U/t=10.5$.
         $\Gamma$, M, and K are defined in Fig.~\ref{cluster.fig}(b).
        }
\label{spectral_function.fig}
\end{figure}

\begin{figure}[t]
\includegraphics[width=8.0cm]{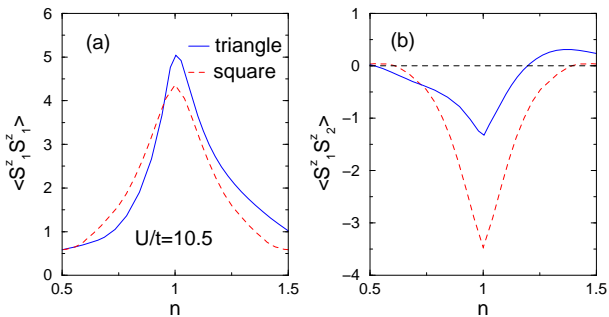}
\caption{(color online) Static (a) local and (b) nearest neighbor spin 
         correlations for $U/t=10.5$ with doping.
         The dashed curves are the corresponding results 
         on a square lattice.
        }
\label{spin.fig}
\end{figure}

   Figure~\ref{spectral_function.fig} shows the spectral function
$A(\vec{k},\omega)$ for both $10\%$ hole- and electron-dopings.  
Electron-doping corresponds to the negative sign of the hopping integral 
($t<0$) in the $t-J$ model.
At half-filling (Fig.~\ref{spectral_function.fig}(b))
$A(\vec{k},\omega)$ shows several features, which are similar to those 
on a square lattice~\cite{KKSTCK:2006}. 
Although they are weaker than on a square lattice, the low energy inner bands   
are clearly seen near the Fermi energy, which are caused by short-range  
spin correlations~\cite{KKSTCK:2006}.
Upon hole- or electron-doping the Fermi level jumps to the nearest 
low energy band just like on a square lattice.
This indicates the importance of 
short-range correlations even on a highly
frustrated triangular lattice.
The most dramatic difference, however, is that near half-filling  
there is no evidence of a (strong coupling) pseudogap
unlike on a square lattice.
It is more evident from the fact that $A(\vec{k},\omega)$ becomes 
sharper as $\vec{k}$ approaches the Fermi surface.
This is true for other fillings ($5-30\%$ hole- and electron-doping).
Thus the present result clearly shows that a doped triangular Mott  
insulator is a correlated Fermi liquid unlike on a square 
lattice~\cite{KKSTCK:2006}.
The absence of the pseudogap is ascribed to too much frustration of 
short-range spin correlations on a triangular lattice as shown 
in Fig.~\ref{spin.fig}.
Here we predict that near half-filling  
$A(\vec{k},\omega)$ has a quasiparticle peak at the Fermi surface, 
which should be tested by angle resolved photoemission spectroscopy (ARPES) 
experiments on a triangular lattice compound such as Na$_{x}$CoO$_{2}$
at small doping.
\begin{figure}[t]
\includegraphics[width=8.0cm]{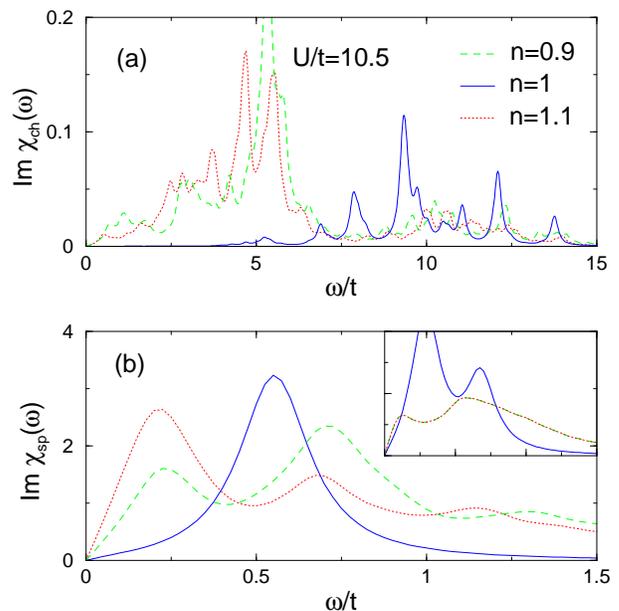}
\caption{(color online) Imaginary part of the local dynamical (a) charge 
         and (b) spin
         correlation functions for $U/t=10.5$ with doping.
         The inset in (b) is the imaginary part of the local dynamical
         spin correlation function for the unfrustrated square lattice case.
         The dashed, solid, and dotted curves correspond to
         $n$=0.9, 1 and 1.1, respectively.
        }
\label{charge_spin_doping.fig}
\end{figure}

   Next let us examine the local dynamical charge and spin
correlation functions upon doping shown in Fig.~\ref{charge_spin_doping.fig}.
As expected, with doping the high energy charge excitations at half-filling
move to the low energy due to the metallic nature of electrons. 
The most surprising result comes from the spin correlation function.
The primary low energy peak near $w/t=0.2$, which has vanished 
in the insulating state due to strong geometric frustration
shown in Fig.~\ref{charge_spin.fig}, \textit{reappears} upon doping.
This is completely opposite to the square lattice case  
(the inset in Fig.~\ref{charge_spin_doping.fig}(b)) where 
the primary low energy peak weakens rapidly with doping.
On a triangular lattice geometric frustration is released by doping.
Here we also predict that inelastic neutron scattering experiments should 
observe the increase of low energy spin excitations with doping  
if the undoped system is a Mott insulator (see caveat~\cite{spin_excitations}).

   Finally we study the static local and nearest neighbor spin 
correlations with doping 
for both triangular and square lattices shown in Fig.~\ref{spin.fig}.
For the local quantity $\langle S^{z}_{1} \cdot S^{z}_{1} \rangle$
they are similar in both cases, although the particle-hole symmetry 
is broken on a triangular lattice.
The most dramatic difference is seen in the nearest neighbor spin
correlation $\langle S^{z}_{1} \cdot S^{z}_{2} \rangle$.
First of all the nearest neighbor spin correlation is much reduced 
compared with a square lattice because of strong geometric frustration.
While it remains essentially antiferromagnetic (negative sign)
on a square lattice, it becomes rapidly \textit{ferromagnetic} 
on the electron-doped side of the triangular lattice.
The strong asymmetric behavior on a triangular lattice that the spin 
correlation becomes ferromagnetic only on the electron-doping side 
is in agreement with 
the high temperature expansion study of the $t-J$ model
by Koretsune \textit{et al.}~\cite{Koretsune:2002} and 
the analysis of an isolated three-site cluster
by Merino \textit{et al.}~\cite{Merino:2005}.
It is also noteworthy that the ferromagnetic correlation is maximum 
near $n=1.35$ ($x=0.35$), where superconductivity has been observed in 
Na$_{x}$CoO$_{2}$ $\cdot$ yH$_{2}$0~\cite{Takada:2003}.
Although the small cluster used in this work ($N_{c}=3$) does not allow   
us to study which gap symmetry (p or f) is more stable  
and how robust this result 
would be with respect to several parameters such as doping
level,
our present result offers a natural route to
\textit{spin triplet} superconductivity in this compound
within the one-band Hubbard model (without assuming the existence of
unobserved hole pockets near $\vec{k}=K$).

   Sodium cobaltate Na$_{x}$CoO$_{2}$ has shown a rich phase 
diagram~\cite{Foo:2004}. These include a superconducting phase  
near $x=0.3-0.35$ upon hydration, a charge-ordered insulating phase
at $x=0.5$ and a spin-density-wave metallic phase above $x=0.75$.
Since the charge-ordered insulating phase in cobaltates is observed
at $x=1/2$ doping which is not a natural commensurate value
($x=1/3$ or $2/3$) of the triangular lattice, the one-band Hubbard  
model alone would not be enough to describe this phase.
It requires longer-range Coulomb interaction $V_{ij}$, which we did not 
consider in this work.

   In conclusion,
we have studied 
electronic properties of the Hubbard model on a triangular lattice
using the cellular dynamical mean-field theory.
The interplay of strong geometric frustration and electron correlations
is responsible for many unconventional features, which are 
in stark contrast to the unfrustrated square lattice case.
In particular the appearance of a Fermi liquid upon doping is ascribed to
too much frustration of short-range spin correlations on a triangular lattice.
We predict that upon doping a triangular Mott insulator
ARPES should observe a quasiparticle peak (no pseudogap) 
at the Fermi surface
and that inelastic neutron scattering should find an increase of low energy 
spin excitations. 

   We thank P. W. Anderson, S. R. Hassan, and A. -M. S. Tremblay for useful 
discussions, and A. -M. S. Tremblay for critical reading of the 
manuscript.
Computations were performed on the Elix2
Beowulf cluster and on the Dell cluster of the RQCHP. The present work was
supported by NSERC (Canada), FQRNT (Qu\'{e}bec), CFI (Canada), and CIAR.



\begin{thebibliography}{99}

\bibitem{Anderson:1973} P. W. Anderson, Mat. Res. Bull. \textbf{8}, 153 (1973);
                        Science \textbf{235}, 1196 (1987).

\bibitem{Takada:2003} K. Takada, H. Sakurai, E. T. Muromachi,
         F. Izumi, R. A. Dilanian, and T. Sasaki,
         Nature (London) \textbf{422}, 53 (2003).

\bibitem{Shimizu:2003} Y. Shimizu, K. Miyagawa, K. Kanoda, M. Maesato,
         and G. Saito,
         Phys. Rev. Lett. \textbf{91}, 107001 (2003).

\bibitem{Koretsune:2002} T. Koretsune and M. Ogata, 
         Phys. Rev. Lett. \textbf{89}, 116401 (2002).

\bibitem{Baskaran:2003} G. Baskaran, 
         Phys. Rev. Lett. \textbf{91}, 097003 (2003).

\bibitem{Kumar:2003} B. Kumar and B. S. Shastry, 
         Phys. Rev. B \textbf{68}, 104508 (2003).

\bibitem{Wang:2004} Q.-H. Wang, D.-H. Lee, and P. A. Lee, 
         Phys. Rev. B \textbf{69}, 092504 (2004).

\bibitem{Honerkamp:2003} C. Honerkamp, 
         Phys. Rev. B \textbf{68}, 104510 (2003).

\bibitem{Capone:2001} M. Capone, L. Capriotti, F. Becca, and S. Caprara, 
         Phys. Rev. B \textbf{63}, 085104 (2001).

\bibitem{Haerter:2005} J. O. Haerter and B. S. Shastry, 
         Phys. Rev. Lett. \textbf{95}, 087202 (2005);
         J. O. Haerter, M. R. Peterson, and B. S. Shastry,
         cond-mat/0608005.

\bibitem{Tohyama:2006} T. Tohyama, 
         cond-mat/0605007.

\bibitem{Bulut:2005} N. Bulut, W. Koshibae, and S. Maekawa, 
         Phys. Rev. Lett. \textbf{95}, 037001 (2005).

\bibitem{Merino:2005} J. Merino, B. J. Powell, and R. H. McKenzie, 
         Phys. Rev. B \textbf{73}, 235107 (2006).

\bibitem{Aryanpour:2006} K. Aryanpour, W. E. Pickett, and R. T. Scalettar, 
         Phys. Rev. B \textbf{74}, 085117 (2006).

\bibitem{Weber:2006} C. Weber, A. L\"{a}uchli, F. Mila, and T. Giamarchi, 
         Phys. Rev. B \textbf{73}, 014519 (2006).

\bibitem{Kotliar:2001} G. Kotliar, S. Savrasov, G. Pallson, and G. Biroli,
         Phys. Rev. Lett. \textbf{87}, 186401 (2001).

\bibitem{GKKR:1996} A. Georges, G. Kotliar, W. Krauth, and M. J. Rozenberg,
         Rev. Mod. Phys. \textbf{68}, 13 (1996).

\bibitem{Bolech:2002} C. Bolech, S. S. Kancharla, and G. Kotliar,
         Phys. Rev. B \textbf{67}, 075110 (2003);
         M. Capone, M. Civelli, S. S. Kancharla, C. Castellani, and G. Kotliar,
         Phys. Rev. B \textbf{69}, 195105 (2004).

\bibitem{KKT:2006} B. Kyung, G. Kotliar, and A. M. S. Tremblay,
         Phys. Rev. B \textbf{73}, 205106 (2006).

\bibitem{KKSTCK:2006} B. Kyung, S. S. Kancharla, D. S\'{e}n\'{e}chal,
         A. -M. S. Tremblay, M. Civelli, and G. Kotliar,
         Phys. Rev. B \textbf{73}, 165114 (2006).

\bibitem{Caffarel:1994} M. Caffarel and W. Krauth,
         Phys. Rev. Lett. \textbf{72}, 1545 (1994).


\bibitem{comment} The position of this peak scales as $J=4t^{2}/U$.

\bibitem{spin_excitations} If the undoped system is a metal, the opposite
         result would occur.


\bibitem{Foo:2004} M. L. Foo, Y. Wang, S. Watauchi, H. W. Zandbergen,
         T. He, R. J. Cava, and N. P. Ong,
         Phys. Rev. Lett. \textbf{92}, 247001 (2004).

\end{thebibliography}
\end{document}